\documentclass[sigconf]{acmart}
\AtBeginDocument{%
  }

\setcopyright{acmlicensed}
\copyrightyear{2026}
\acmYear{2026}
\acmDOI{XXXXXXX.XXXXXXX}
\acmConference[AIDA2J at ICAIL '26]{Workshop on Artificial Intelligence for Access to Justice, at the 21st International Conference on Artificial Intelligence and Law}{June 08--12,
  2018}{SMU, Singapore}
\acmBooktitle{Preprints for AIDA2J 2026}

\usepackage{makecell}
\usepackage{algpseudocode}
\usepackage{algorithm}
\usepackage{tikz}
\usetikzlibrary{arrows.meta}
\begin{document}

\title[Explainable and Accurate Conversational Agents for eGovernment]{
Hybrid AI for Explainable and Accurate\\ Conversational Agents in eGovernment}


\author{Ilias Chalkidis}
\email{ilias.chalkidis@di.ku.dk}
\orcid{}
\affiliation{%
  \institution{Copenhagen University}
  \city{Copenhagen}
  \country{Denmark}
}

\author{Vlad Paul Cosma}
\email{vco@di.ku.dk}
\orcid{}
\affiliation{%
  \institution{Copenhagen University}
  \city{Copenhagen}
  \country{Denmark}
}
\author{Søren Debois}
\email{debois@dcrsolutions.net}
\orcid{}
\affiliation{%
  \institution{DCR Solutions}
  \city{Copenhagen}
  \country{Denmark}
}

\author{Daniel Hershcovich}
\email{dh@di.ku.dk}
\orcid{}
\affiliation{%
  \institution{Copenhagen University}
  \city{Copenhagen}
  \country{Denmark}
}

\author{Thomas Hildebrandt}
\email{hilde@di.ku.dk}
\orcid{}
\affiliation{%
  \institution{Copenhagen University}
  \city{Copenhagen}
  \country{Denmark}
}

\author{Hugo A. L\`opez}
\email{hulo@dtu.dk}
\orcid{}
\affiliation{%
  \institution{Technical University of Denmark}
  \city{Kgs. Lyngby}
  \country{Denmark}
}

\author{Amogh Raina}
\email{amogh.raina@di.ku.dk}
\orcid{}
\affiliation{%
  \institution{Copenhagen University}
  \city{Copenhagen}
  \country{Denmark}
}

\author{Konstantinos Varvoutas}
\email{konva@dtu.dk}
\orcid{}
\affiliation{%
  \institution{Technical University of Denmark}
  \city{Kgs. Lyngby}
  \country{Denmark}
}

\author{Tilman Zuckmantel}
\email{tizu@di.ku.dk}
\orcid{}
\affiliation{%
  \institution{Copenhagen University}
  \city{Copenhagen}
  \country{Denmark}
}

\renewcommand{\shortauthors}{Hildebrandt et al.}

\begin{abstract}
 We present a so-called Conversational Hybrid AI (CHAI) architecture for building explainable and accurate 
  conversational agents for eGovernment. We exemplify the architecture with a running prototype of a Covid-19 Chatbot based on a governmental guideline directed to citizens. We also describe an ongoing case on case management for supplementary grants for students with disabilities. We use large language models (LLMs) as a bounded conversational
interface to a rule-based (symbolic AI) controller that executes a logical model expressing the provisions and obligations of the law and/or guidelines. As logical modelling language we use
 Dynamic Condition Response (DCR) graphs, a symbolic
declarative process-modeling language developed with the aim to be able to express both deontic, defeasible and temporal logic properties, making it suitable for expressing both the rules of the law and the steps of the legal case management processes. 
 
\end{abstract}


\maketitle

\section{Introduction}

Recent advances in the development of large language models (LLMs) and the rollout of commercial LLM-based chatbot assistants based on the retrieval augmented generation (RAG) architcture have raised high expectations for the potential impact of such technologies across industries.
However, several critical concerns have been raised about the trustworthiness of end-to-end LLM-based RAG solutions~\cite{https://doi.org/10.1111/jels.12413}. Such concerns dictate the conversational framework as a whole, from the interpretation of information provided to the models as user input (prompt), the retrieval of appropriate information from its parametric (compressed in the model parameters)  and non-parametric (externally retrieved documents) memory, the execution of tasks based on both, and the dissemination of decisions and results in the form of natural text (response). While in many use cases, e.g., Q\&A of common knowledge, code development assistance, end-to-end LLM-based solutions can be quite effective, if the user is aware of and ready to cope with the possible incorrect outputs, the situation is quite different in high-stakes scenarios, such as those commonly encountered in processes that involve the interpretation and execution of law, i.e., case management in the public sector.

In this work, we propose an architecture for developing transparent, accurate and explainable conversational user interfaces for eGovernment. Concretely, we propose treating LLMs as a bounded conversational interface to a rule-based \emph{controller} (similar to the proposal in~\cite{belcak2025smalllanguagemodelsfuture}) executing logical rules expressed as Dynamic Condition Response (DCR) graphs~\cite{hildebrandt2011declarative,hildebrandt2021decision}.
As illustrative examples we show a conversational agent for guiding citizens who have been in close contact with a person infected with Covid-19 and also show a small part of the coding of an official guideline for
handling applications for study grants for students with disabilities.
The work presented in the paper is part of a national project \emph{Explainable Hybrid AI for Computational Law and Accurate Legal Chatbots} (XHAILE) supported by Innovation Fund Denmark.\footnote{https://di.ku.dk/english/research/research-projects/xhaile/}.

\section{Background}

\subsection{The Promise and Peril of End-to-End LLMs}
The application of LLMs to the legal domain has accelerated rapidly, driven by their ability to process unstructured legal text, draft documents, and apparently answer queries with high fluency. The news that such general-purpose models seemed to outperform specialized models on bar exams and statutory reasoning tasks \cite{katz2024gpt4,openai2024gpt4} spread rapidly around the world.\footnote{The bar exam result was later questioned~\cite{BarExam2024}.} Consequently, there has been a push to explore if LLMs could be used to 
provide digital support to citizens as part of eGovernment services, e.g. by answering questions to the law. Many approaches employ a
Retrieval Augmented Generation (RAG) architecture~\cite{lewis2021retrievalaugmentedgenerationknowledgeintensivenlp}, where a chatbot first searches in a curated database containing up-to-date law texts and guidelines and indexed by a vector embedding. The search results are then appended to the question and used as prompt to the LLM, which generates the answer. 

However, the probabilistic nature of the LLMs presents a fundamental barrier to adoption in high-stakes public sector case management. LLMs are prone to ``hallucinations''~\cite{https://doi.org/10.1111/jels.12413} and generating plausible but factually incorrect legal citations or reasoning paths \cite{dahl2024largelegal}. In legal contexts, where ``correctness'' is defined by strict adherence to statute and procedure, a probabilistic approximation of the law is insufficient. Furthermore, the ``black box'' nature of artificial neural networks creates an explainability gap; an LLM cannot provide a verifiable trace of its reasoning, making it impossible to audit whether a decision was derived from the correct provision or merely synthesized from training data \cite{stanford2025legaldesign}. The combination of a technology that provides factually incorrect answers by design and the lack of explainability also constitutes a problem for accountability, since a municipality or case worker cannot reasonably be held accountable for an answer from a chatbot, if they are not presented with a valid  legal reasoning supporting it.

These limitations push us toward architectures where LLMs never act as autonomous decision-makers facing citizens, but instead serve as interfaces to symbolic, knowledge-based models, that are based on explicit legal rules and well-defined methods for inference of conclusions.

\subsection{Symbolic Legal Reasoning}
Before the LLM era, legal AI focused on symbolic approaches, encoding laws as explicit rules using formalisms like Logic Programming (e.g., Prolog, ASP) or semantic web ontologies \cite{benchcapon2020legalai,satoh2010prolog}. These systems offer what LLMs lack: perfect consistency, auditable provenance, and guaranteed adherence to encoded constraints.

However, purely symbolic systems have historically failed to scale due to the ``knowledge acquisition bottleneck.'' Translating natural language statutes into formal code requires immense manual effort and often results in brittle systems that break when faced with ambiguous user input or edge cases not explicitly modeled \cite{susskind2019onlinecourts}. Moreover, traditional logic engines (e.g., SMT solvers) typically represent the law as code that is difficult to understand as a whole by legal experts and treat the law as a set of static constraints to be satisfied. While powerful for checking consistency, they are less natural for modeling the temporal and defeasible logical~\cite{defeasible} lifecycle of a case (e.g., ``Step A must happen before Step B, but only if deadline C has not passed''). 

\subsection{The Neuro-Symbolic Convergence}
To bridge the divide between artificial neural networks and symbolic models, the field is shifting toward \textit{Neuro-Symbolic AI}, which we will refer to as \textit{hybrid AI}, which combines statistical associations represented in the sub-symbolic parameters of artificial neural networks and learned from data with the flexibility and logical rigor of symbolic systems \cite{garcez2023neurosymbolic,belle2025relevance}. 
There are generally two overall approaches to Hybrid-AI. In the first approach, the symbolic systems are used to feed, monitor, and guard the sub-symbolic, artificial neural network models, which, however, still carry out the inference of the answer. In the second approach, the sub-symbolic models are used to feed, e.g., as translators of inputs from users or knowledge bases, the symbolic models, which carry out the inference of the answer.

In the legal domain, recent frameworks such as \textit{LawGiBa}~\cite{lawgiba2024} are using the second approach, 
to convert natural language queries into formal logic (e.g., translating a tax question into a Prolog query) ~\cite{pan2023logiclm, lawgiba2024}. 
In this architecture, the LLM handles the ``unstructured'' world (interpreting user intent, extracting facts), while a symbolic solver handles the ``structured'' world (executing the rules). We will also follow this approach, since the grounding in formal symbolic reasoning ensures that the final output is not a hallucination, but a natural language description of a valid logical proof derived by the symbolic engine. Yet, the challenge of modeling \emph{dynamic} administrative processes and defeasible logical rules as exemplified above remains distinct from static statutory interpretations.

\subsection{From Static Logic to Dynamic Process Logic}
\label{sec:DPL}
While neuro-symbolic approaches are promising, we argue that the interaction between citizens and governmental institutions is not merely a static logical one. Public administration is governed by a mix of deadlines, sequences of required actions, and permissions and obligations that may change over time as the case unfolds. This calls for a combination of decision, temporal, and defeasible rules.

\begin{figure*}[h]
  \centering
  \includegraphics[width=\textwidth]{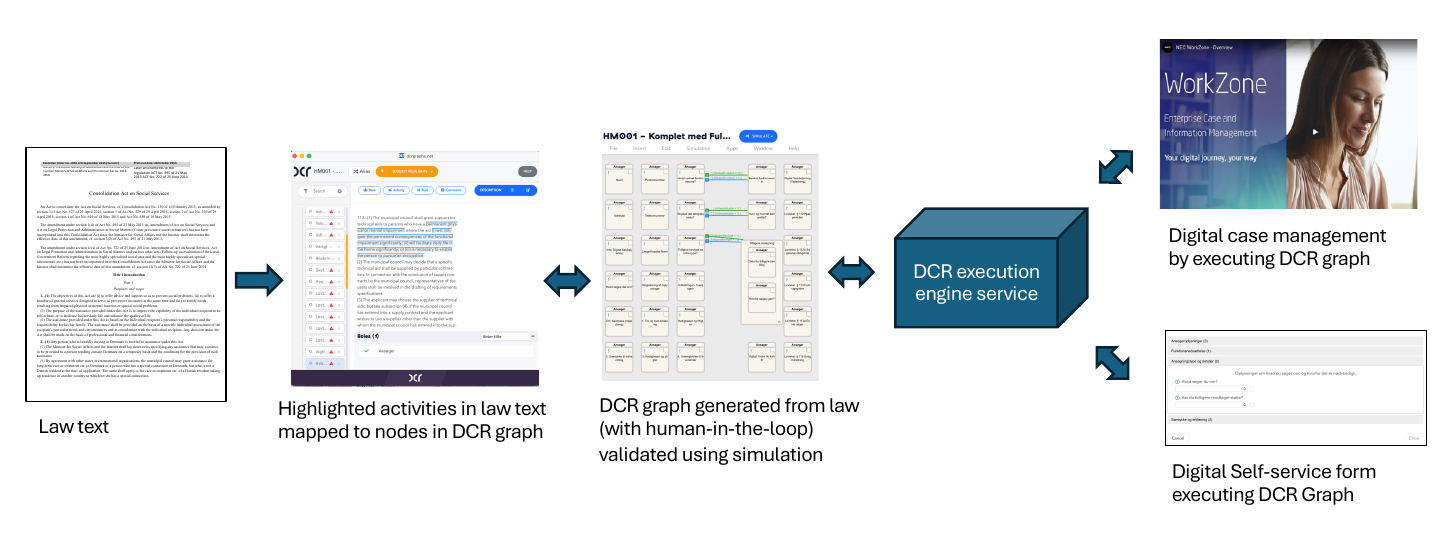}
  \caption{Rules and activities in law texts can be highlighted by domain-experts and thereby translated to a DCR graph in the DCR no-code design tool (dcrsolutions.net), that can be simulated for validation and executed in a case management processes and digital self-service forms using the DCR execution engine service.}
  \Description{Infographic showing the AI-assisted (AI with human-in-the-loop) translation of law texts to DCR graphs that can be executed as digital self-service solutions and case management processes.}
  \label{fig:eGovPresent}
\end{figure*}

Specifically, we advocate for \textbf{Dynamic Condition Response (DCR) graphs}, a declarative executable process modelling language with support for deontic and defeasible temporal constraints \cite{hildebrandt2011declarative}. Unlike imperative process modelling notations such as BPMN \cite{OMG.BPMN}, that rigidly describe the sequencing of possible steps as flow charts, DCR graphs model the rules defining only what is forbidden,  or required, leaving everything else allowed. This aligns naturally with the nature of law, which often prescribes  prohibitions and obligations, and proscribes conduct, rather than scripting it. 
The modelling language of DCR graphs was introduced in~\cite{hildebrandt2011declarative}. Since then, the language has been extended to be able to model processes with time~\cite{DCRtime}, data and decisions~\cite{hildebrandt2021decision}, introducing data-dependent rules and embedded computations, and also integrating the Decision Model and Notation (DMN) standard~\cite{DMN}.
The task of modelling is  also supported by commercial cloud-based solutions, including
tools for simulation of processes and for assisting domain and legal experts in the mapping of text based process models, guidelines, and regulations into DCR graphs~\cite{2caa2d974cdd436ebb27ee81f4e91251}.
DCR graphs have been successfully deployed in Danish public-sector case-management systems using a \emph{process engine web service}, that is integrated into commercial platforms~\cite{marquard2017dcr}. Fig.~\ref{fig:eGovPresent} illustrates the use of DCR graphs as computational law. By using the highlighter mapping tool, a legal expert interacts with an NLP tool to highlight and translate the legal text onto a DCR graph~\cite{lopez2019assisted}. Using the DCR execution engine, the DCR graph can be validated in the design tool and also be executed as a self-service form and in a case management tool such as WorkZone, thereby handling the case management process after the form has been submitted.

However, after the introduction of publicly available chatbots that can interact fluently with users in natural language, there has been a push for introducing chatbots instead of (or in addition to) the classical interfacing with eGovernment services using web forms.
Recent work has begun exploring the combination of the sub-symbolic LLMs and  declarative, symbolic rules models, demonstrating that LLMs can effectively interpret user intents for the modelling and discovering of declarative models~\cite{grohs2025declarative,debois2024llmui,lindner2025discovering}.

In the XHAILe research project we aim to utilize (extensions of) DCR Graphs as symbolic models of the law that can be used to accurately control the answers of conversational interfaces between citizens and the government. 
The graphs should first support and ultimately replace the current electronic guidelines and form filling, and second, support the subsequent case management, including feedback interactions, to result in an answer provided to the citizen, with possible enactments of the decisions involved in the process. 
\begin{figure}
  \centering
\includegraphics[width=.5\textwidth]{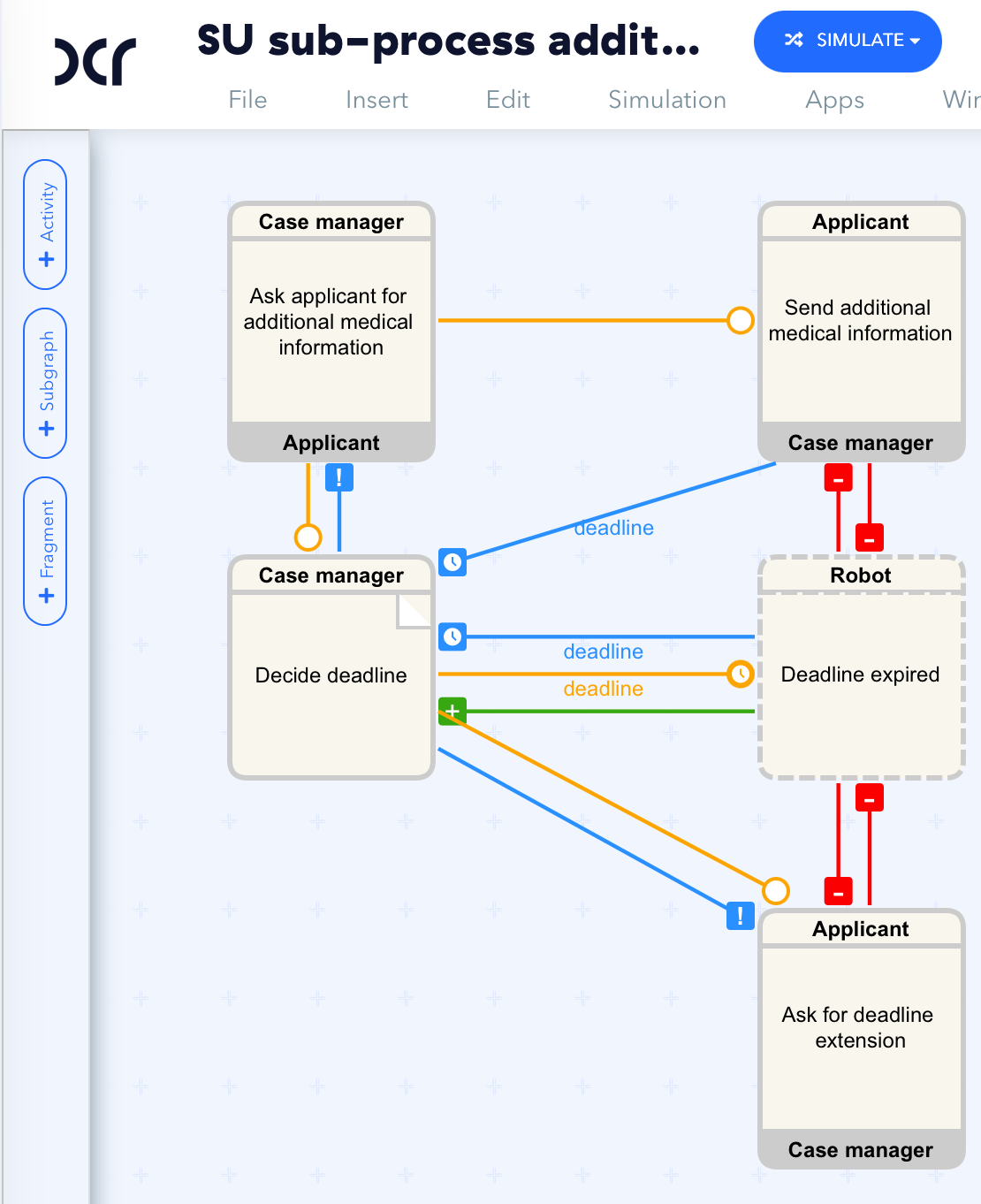}
  \caption{A DCR Graph coding a sub process of the handling of Study Grants for students with dissabilities.}
  \Description{A DCR Graph coding a sub process in the handling of Study Grants for students with dissabilities.}
  \label{fig:SUsubprocess}
\end{figure}

\begin{figure}
  \centering
\includegraphics[width=.5\textwidth]{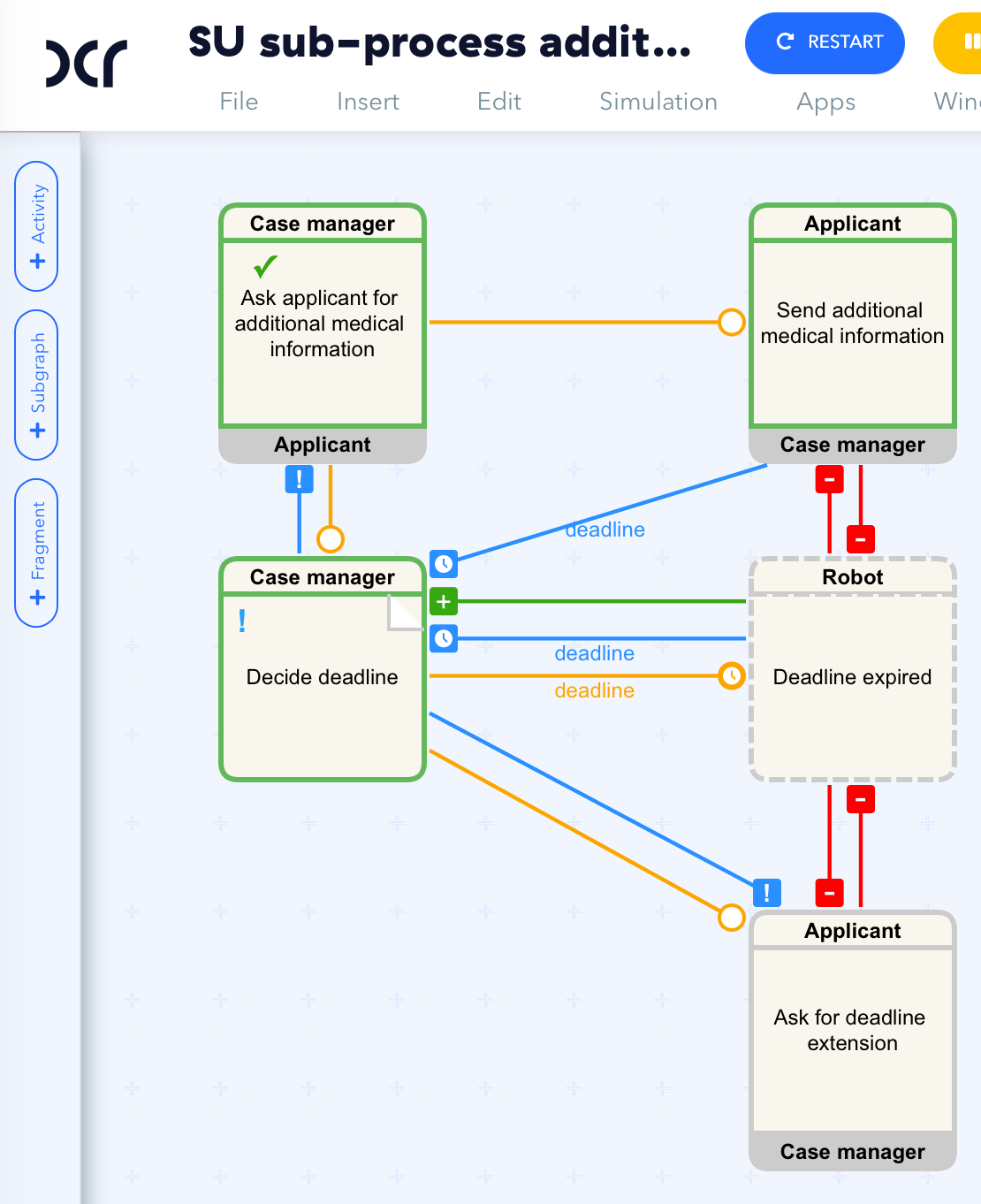}
  \caption{The DCR graph with updated marking after \texttt{Ask applicant for additional medical information} has been executed (shown by a checkmark in the top left corner) and \texttt{Decide deadline} is a pending obligation (shown by an exclamation mark in the top left corner).}
  \Description{DCR graph with updated marking after \texttt{Ask applicant for additional medical information} has been executed.}
  \label{fig:SUsubprocessexecuted}
\end{figure}

\begin{figure}
  \centering
\includegraphics[width=.5\textwidth]{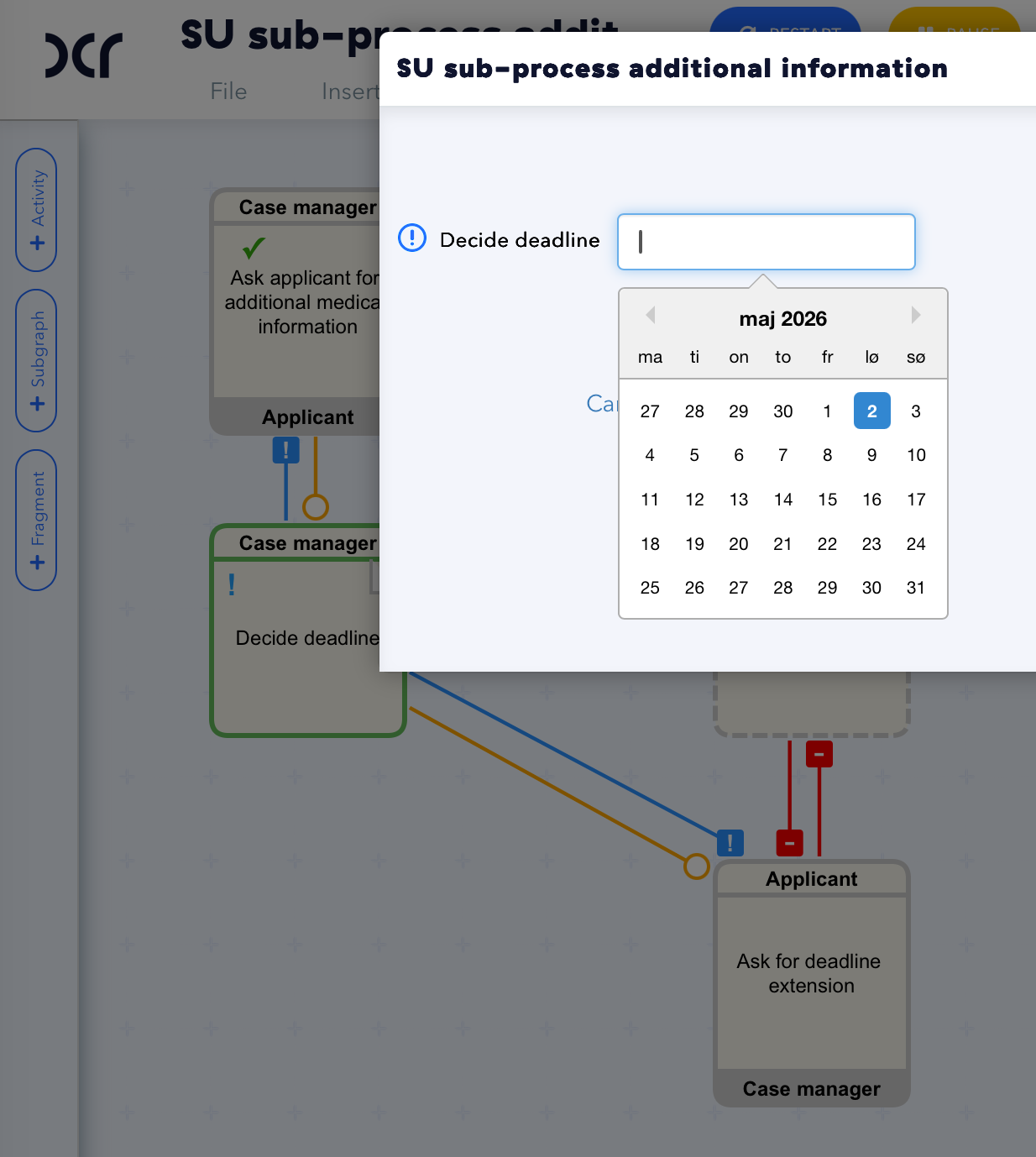}
  \caption{Selecting the deadline in the simulation tool.}
  \Description{A figure showing how the deadline activity is executed in the simulation tool.}
  \label{fig:decidedeadline}
\end{figure}

\begin{figure}
  \centering
\includegraphics[width=.5\textwidth]{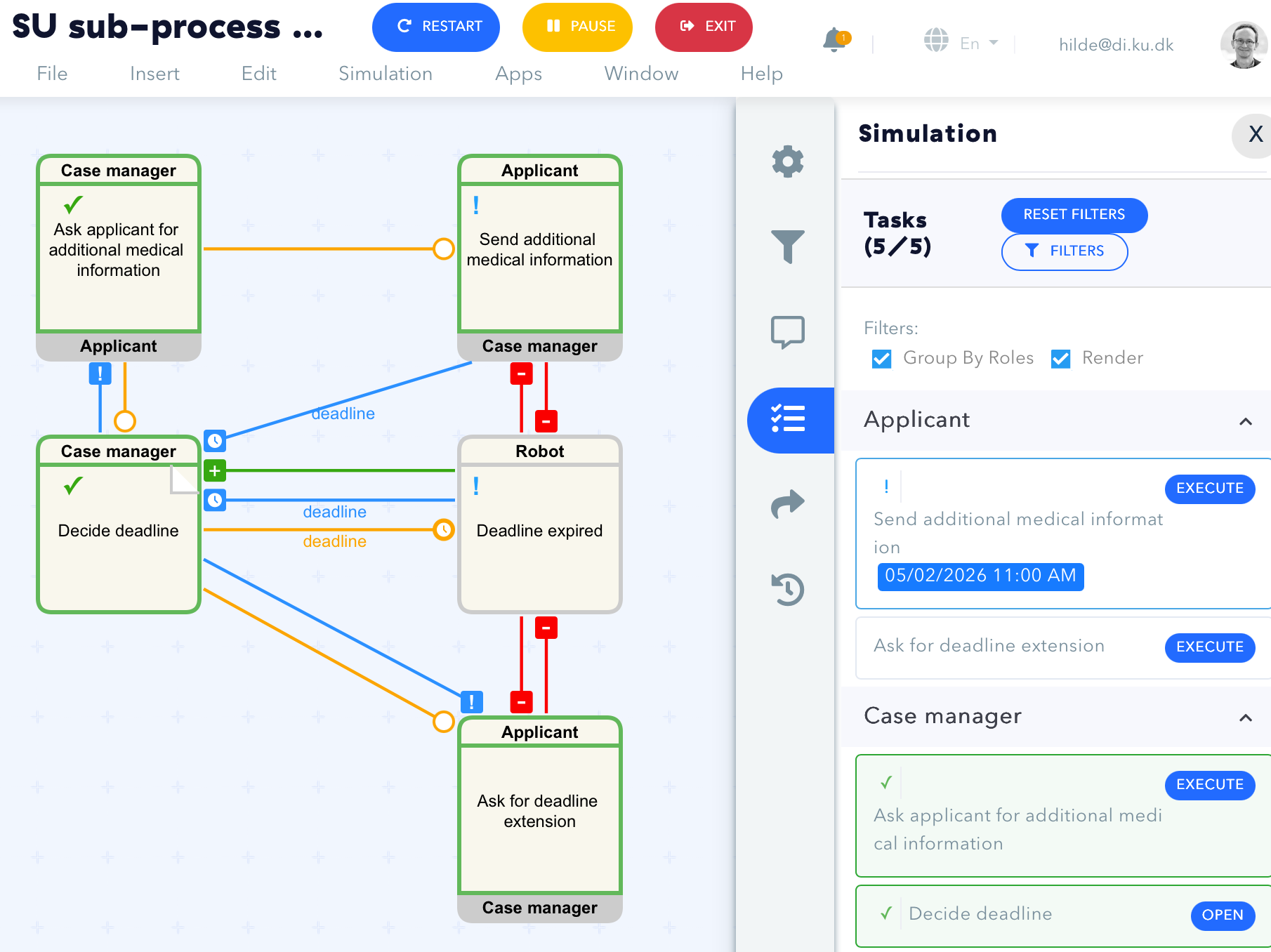}
  \caption{After selecting the deadline, the activities \texttt{Send additional medical information} and \texttt{Deadline expired} are both pending obligations within the deadline, but \texttt{Deadline expired} is not enabled before the deadline has been reached, because the condition to \texttt{Decide deadline} has a delay.}
  \Description{A figure showing the state of the graph after the deadline has been decided.}
  \label{fig:decidedeadline2}
\end{figure}

\section{Example Guidelines}
We consider the digitalization of two different guidelines. The first guideline, briefly described in Sec.~\ref{sec:covid19}, was published by the danish government during the Covid-19 pandemic and is directed towards citizens. The guideline describes what to do in case of close contact with another person who is infected by Covid-19. The second guideline, briefly described in Sec.~\ref{sec:SU}, describes what a case manager and a citizen should do when the citizen apply for supplementary study grants due to disabilities.
\subsection{Covid-19 Close Contact}
The Covid-19 guidelines for what citizens should do in case of close contact with another person who is infected were frequently updated during the Covid-19 pandemic and distributed as online documents. They are no longer available from the governmental agencies, but examples can still be found e.g. at the websites of some highschools\footnote{https://oerestadgym.dk/wp-content/uploads/2021/04/Naer-kontakt-Corona-.pdf}. The guideline was essentially a decision tree based on questions like wether the citizen live with the infected person, and if not, have been in physical contact with the person and when the last time of contact was, as also shown in Fig.~\ref{fig:CoronaChat}.

Even though the guideline was written to be understandable by citizens, many felt uncertain about how to interpret the guideline and called on the telephone help line and sometimes waiting in a queue for hours. A DCR graph encoding the 11 page guideline can be found at the URL: \url{https://dcrgraphs.net/Tool?id=2001183}. 
\label{sec:covid19}
\subsection{Grants for Students with Disabilities}
The guideline for Grants for Students with Disabilities is published at the official website for danish law, retsinformation\footnote{https://www.retsinformation.dk/eli/retsinfo/2025/9577}. It consists of 21 pages describing both the permissions and obligations of citizens and the government case worker during the handling of an application. For instance, the sub process described as a DCR graph in Fig.~\ref{fig:SUsubprocess} is a coding of subsections 5.2 and 5.3, which describe that the agency may decide to ask the applicant for additional information and in doing that, they must give a deadline. It is written, that the citizen may ask for extension of the deadline, but if the citizen does not ask for extension and does not provide the additional information, then the agency will decide on the basis of available information, which often means that the application will be rejected. In other words, these two sections written on page 14 of a 21 page long document contain important information for the applicant, i.e. that not reacting on a request for additional information will likely lead to rejection.
\label{sec:SU}
\section{Conversational Hybrid AI (CHAI)}
In this section we will present our architecture and prototype for Conversational Hybrid AI (CHAI) combining rule based, symbolic AI based on DCR graphs and sub-symbolic AI based on large language models. We first in Sec.~\ref{sec:DCRintro} give a brief introduction to DCR Graphs explained using a sub process of the Study Grant process as example. We then in Sec.\ref{sec:API} briefly describe a subset of the Application Programming Interface (API) for managing and executing DCR graphs. In Sec.~\ref{sec:Symbol}, Sec.~\ref{sec:LLM} and Sec.~\ref{sec:hybrid} we then describe respectively the purely symbolic, purely sub-symbolic and the hybrid approach with pseudo-code for the implementation.

\subsection{DCR Graphs by Example}
\label{sec:DCRintro}
At its core, a DCR graph describes a process declaratively as a graph where the nodes of the graph represent activities or events, and the edges of the graph represent logical constraints or reactions between the activities/events. 

 A key novel aspect of DCR graphs is that they can also express the intermediate states of a process, i.e. what has already been observed, such as established facts or actions done in the management of a case, what is currently included or excluded from the logical rules, and what needs to happen/be established in the future. 

Moreover, the DCR graphs notation is equipped with a formal semantics that allows to computationally decide which activities can be done at a current state and update the state according to the rules of the graph, if the activity/observation is carried out.

Figure~\ref{fig:SUsubprocess} shows an example of a DCR graph\footnote{in the design tool available at \url{https://dcrsolutions.net}} coding a sub process in the handling of Study Grants for students with disabilities. The graph has five activities, depicted as boxes and constituting the nodes of the graph. Each activity has activity label in the middle (e.g. \texttt{Decide deadline} and \texttt{Ask for deadline extension}) and a role at the top (e.g. \texttt{Case manager} and \texttt{Applicant}) describing who can execute the activity. An activity may also have a role written on dark background at the bottom, defining the role for who gets notified about the execution of the activity. For instance, the activity labeled \texttt{Ask for deadline extension} is carried out by the applicant and the case manager is notified that it is executed. 

The relations between the activities, constituting the edges of the graph, define the logical conditions and reactions for activities. The graph showcases the four core relations of DCR graphs. The (orange) relation with the hollow circle at the one end, for instance between the activity \texttt{Decide deadline} and the activity \texttt{Ask for deadline extension} is called a condition relation and denotes, that the activity at the end with the circle, in this case \texttt{Ask for deadline extension} has to be executed before the other activity, in this case  \texttt{Decide deadline}, 
can be executed. 
Similarly, the condition between \texttt{Ask applicant for additional medical information} and \texttt{Decide deadline} define that the case manager must ask for additional medical information before deciding on the deadline. Since 
\texttt{Ask applicant for additional medical information} is the only activity without a condition, it is the only activity that can be executed in the initial state, where no other activities have been executed so far.
The (blue) relations with a box at one end containing an exclamation mark, define that when the activity at the end of the box gets executed, it becomes an obligation eventually to execute the other activity. For instance, if the applicant ask for deadline extension, it becomes an obligation for the case manager eventually to decide a deadline again. (The case manager may in this case decide to give the same deadline, which amounts to not extending the deadline). Also, if the case manger asks the applicant for additional medical information it not only becomes possible to decide a deadline, it also becomes an obligation. 
A key feature of DCR graphs is that the nodes (activities) also carry information about the relevant information about the process execution so far, which is referred to as the \emph{marking} of the graph. The marking records for each activity wether the activity 1) has been executed (and the time of the last execution), 2) is obligatory to execute in the future (possibly within a deadline), and 3) is currently included in, or excluded from, the graph.
Excluded activities are depicted with a dashed border, as in this case, the activity \texttt{Deadline expired}. Excluded activities are ignored when evaluating the rules of the graph and cannot be executed. Fig.~\ref{fig:SUsubprocessexecuted} shows the marking of the graph after executing the activity \texttt{Ask applicant for additional medical information} as shown in the online simulation tool. Note that now three activites (shown with green border) are enabled for execution: The activity \texttt{Ask applicant for additional medical information} may be executed again, but also the activities \texttt{Decide deadline} and \texttt{Send additional medical information} are now enabled for execution. 

The activity \texttt{Decide deadline} is different than the others by being a (typed) data input activity. That is, when it is executed, a data value must be provided. In this case, the data is of type date and the value provided is the deadline, which is stored in the marking of the graph as an attribute (referred to by the id \texttt{deadline}) of the node \texttt{Decide deadline}. The execution of the activity in the simulation tool is shown in Fig.~\ref{fig:decidedeadline} and the resulting graph is shown in  Fig.~\ref{fig:decidedeadline2}. It is worth noting, that after executing \texttt{Decide deadline}, the activity \texttt{Deadline expired} becomes included (shown with a solid border) and a pending obligation (an exclamation mark in the top left corner). The inclusion is due to the (green) relation with a plus inside the square between \texttt{Decide deadline} and \texttt{Deadline expired}. The pending obligation is due to the (blue) relation with a box and a small clock inside. This relation is a response relation with a deadline, and in this case, the deadline is the one provided by the user when executing the \texttt{Decide deadline} activity.

The activity \texttt{Deadline expired} has the role \texttt{Robot}, which means that it is an activity that is automatically executed by the engine if it is enabled and a pending obligation. However, note that the orange condition relation between \texttt{Decide deadline} and \texttt{Deadline expired} also has a little clock symbol inside the circle, and the relation also has the word deadline on top of it. This means that the condition requires \texttt{Decide deadline} to be executed and delay at least until \texttt{deadline} has been reached, before \texttt{Deadline expired} can be executed. Consequently, the robot will execute \texttt{Deadline expired} exactly when \texttt{deadline} has been reached. The (red) relations with a minus sign inside a square denotes that if \texttt{Deadline expired} is executed, then \texttt{Send additional medical information} or \texttt{Ask for deadline extension}  gets excluded and cannot anymore be executed, and vice versa, if the Applicant executes either \texttt{Send additional medical information} or \texttt{Ask for deadline extension} before \texttt{Deadline expired} is executed by the engine, then the (red) relations with a minus sign inside a square denotes that \texttt{Deadline expired} again gets excluded (and is then not executed by the engine.)

\subsection{API for DCR as a Service}
\label{sec:API}
\begin{figure}
  \centering
\includegraphics[width=.5\textwidth]{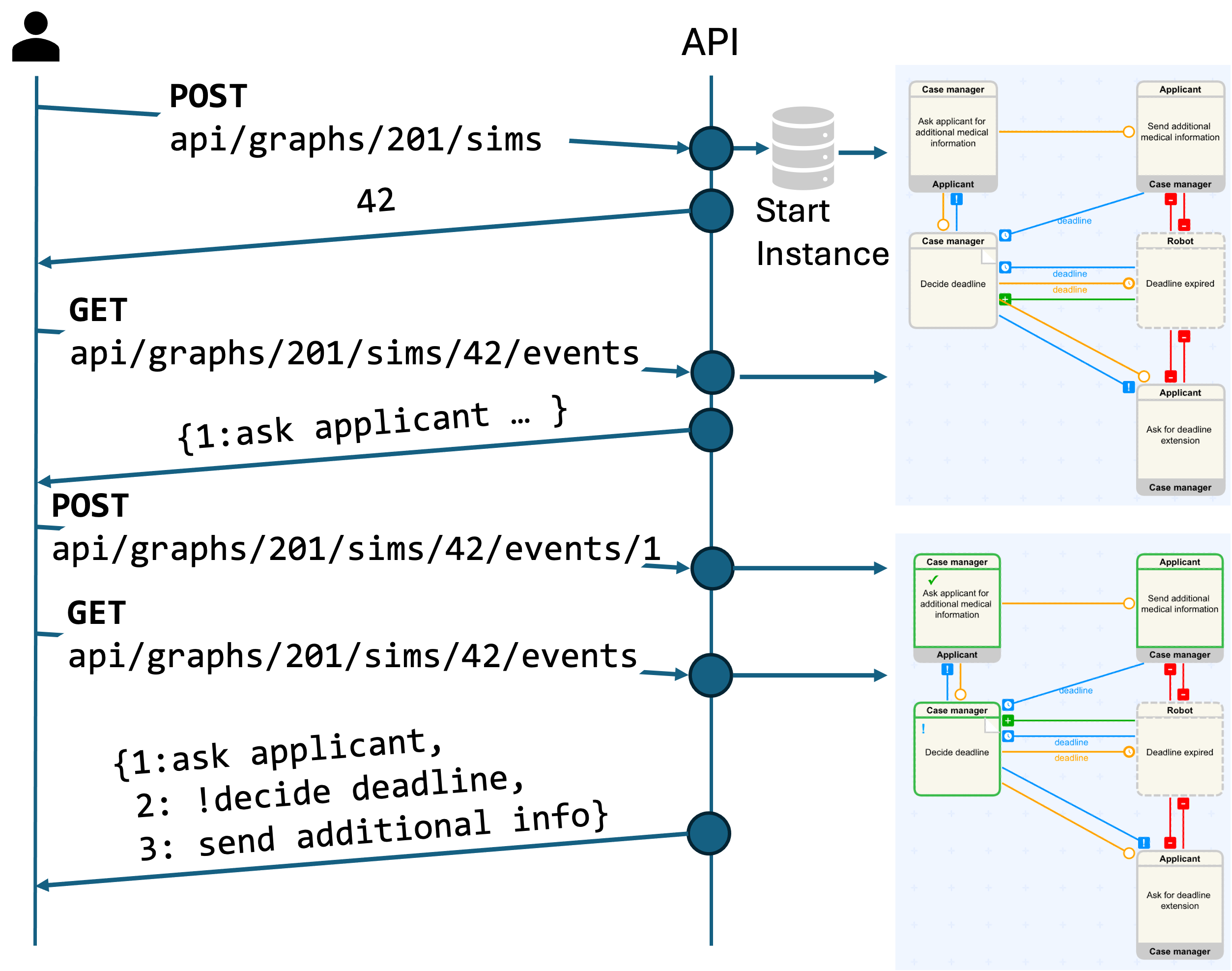}
  \caption{Example interaction between client and API showing the instantiation of an instance of a DCR graph and the execution of the activitiy \texttt{Ask applicant for additional medical information}.}
  \Description{Figure showing the interaction with the API.}
  \label{fig:API}
\end{figure}

To support programmatic process execution using DCR graphs, \textit{DCR Solutions}
offers a RESTful API exposed through the \textit{DCR Active Repository}~\cite{DCRApi}.
The API provides functionality for retrieving and managing graph models, including both the creation and editing of process graphs but also their execution and analysis. In the present paper we focus on the subset of the API for execution of process graphs, which we will refer to as the execution API. This API involves creating and
querying execution instances (referred to as simulation instances or just simulations) and executing enabled activities within a simulation instance. 

Graph models are stored in the repository and can be retrieved and managed through
dedicated endpoints. To illustrate the concepts of user-API interaction, we show in
Fig.~\ref{fig:API} the instantiation of the graph from Fig.~\ref{fig:SUsubprocess}
and the subsequent evolution of this instance through the execution of an activity.
First, the user instantiates a new instance of the DCR graph with identifier~201 by
issuing a \texttt{POST} request. The server starts a new instance and responds with
the instance identifier~42. The user then proceeds to query the current state of the
instance's activities via a \texttt{GET} request, which returns a list of all activities
and their current state. Next, the user executes the \texttt{Ask applicant for additional
medical information} activity by issuing a \texttt{POST} request. An activity is
executed by posting its identifier along with an optional data value, upon which the
engine updates the instance state accordingly. Finally, the user retrieves the updated
activity state, now reflecting that activity~2 (\texttt{Decide deadline}) has become
pending as a result of the execution. Once a simulation instance is complete,
the API further provides access to ordered execution logs and final activity snapshots.
The endpoints used in this paper are
summarized in Table~\ref{tab:dcrapi}.

\begin{table}[htb!]
    \centering
    \small
    \begin{tabular}{l l}
        \toprule
        \textbf{Method \& URI} & \textbf{Description} \\
        \midrule
        \multicolumn{2}{l}{\textbf{Graph Management}} \\
        \texttt{GET /api/graphs/\{id\}} & Retrieve graph def. \\
        \midrule
        \multicolumn{2}{l}{\textbf{Simulation Lifecycle}} \\
        \texttt{POST }$\underbrace{\mathtt{/api/graphs/\{id\}/sims}}_{\mathit{base}}$ & Create sim. instance \\[6pt]
        \texttt{GET } $\mathit{base}$\texttt{/\{simId\}/events} & Get events and state \\
        \texttt{POST } $\mathit{base}$\texttt{/\{simId\}/events/\{eID\}} & Execute an activity \\
        \texttt{GET } $\mathit{base}$\texttt{/\{simId\}/log} & Get execution log \\
        \bottomrule
    \end{tabular}
    \caption{DCR Active Repository API endpoints used.}
    \label{tab:dcrapi}
\end{table}bbb
An important point of having an openly described API for management of DCR graphs and execution of processes is that one can plug in other implementations of rule engines. Concretely, in addition to the DCR design tool and engine used for the present paper, multiple open source DCR tools, such as DCR-JS and DCR4Py ~\cite{Hermansen2024, christfort2025dcr}, exist. 

\subsection{Purely Symbolic Interfaces}
\label{sec:Symbol}

Rule-based conversational interfaces present questions as fixed interface elements and advance through a predefined process flow according to selected values. Every input is validated at entry, every branch is a predictable consequence of prior answers, and the
decision is fully traceable, where no interpretation step could
introduce error~\cite{benchcapon2020legalai}. However, the user is required to map their own situation onto the system's categorical vocabulary before any input is possible, with no opportunity to express themselves in natural language or seek clarification from the interface. This imposes significant cognitive load, particularly for users with limited administrative literacy~\cite{trends-egov-chatbots}, and purely structured interfaces
have been criticized in eGovernment studies for feeling like ``filling out a form'' technically complete but communicatively impoverished~\cite{public-value-chatbots}.

\begin{algorithm}[t]
\caption{Purely symbolic conversational interface driven by a rule-based process engine.}
\label{alg:symbolic}
\begin{algorithmic}[1]
\Require Process graph $G$
\State $sim \gets \textsc{CreateSimulation}(G)$
\State $history \gets \emptyset$
\State $enabled \gets sim.\textsc{GetEnabledEvents}()$
\State $pending \gets sim.\textsc{GetPendingEvents}()$
\While{$pending \neq \emptyset$}
  \If{$enabled \cap pending \neq \emptyset$}
    \State $e \gets \textsc{First}(enabled \cap pending)$
  \Else
    \State $e \gets \textsc{First}(enabled)$ \Comment{pending events exist but none are enabled yet}
  \EndIf
  \State present $\textsc{Question}(e)$ and $\textsc{Options}(e)$ to user
  \State $value \gets$ user selects one of $\textsc{Options}(e)$
  \State $sim.\textsc{Execute}(e, value)$
  \State $history \gets history \cup \{(e, value)\}$
  \State $enabled \gets sim.\textsc{GetEnabledEvents}()$
  \State $pending \gets sim.\textsc{GetPendingEvents}()$
\EndWhile
\State \Return $sim.\textsc{Conclusion}()$
\end{algorithmic}
\end{algorithm}

Algorithm~\ref{alg:symbolic} formalizes this interaction. The interface is driven by the
rule-based DCR process engine that maintains the authoritative state of the conversation: at any point, $sim$ exposes the set of activities that are currently enabled (whose
preconditions are satisfied), pending and included (required to occur)
in the current state). Both sets are fetched explicitly at the start of each iteration
and refreshed after every execution. At each turn, the system selects the first event
from the intersection of enabled and pending activities, where possible; if pending
events exist, but none are currently enabled, meaning their preconditions have not yet
been satisfied, the system instead selects the first enabled event, allowing the user
to make progress that may unlock the pending ones. The engine then executes the selected
event with the chosen value, recomputing the enabled, pending, and excluded sets according to the graph's constraint semantics. The loop terminates when no events remain pending, at which point the engine returns its authoritative conclusion.\footnote{A DCR graph may deadlock, if some event remain pending but no events are enabled. Core and timed DCR graphs with bounded deadlines and delays can be mapped to finite state Büchi-automata that can be analysed for deadlock freedom~\cite{hildebrandt2011declarative,DCRtime}.} There is no interpretation step, and the user has no mechanism to express anything that the predefined options do not anticipate.

\subsection{Purely LLM-Based Approaches}
\label{sec:LLM}

Retrieval-Augmented Generation (RAG) architectures have become the dominant approach for building natural language administrative assistants: a retrieval component surfaces relevant passages from a document corpus, which are provided as context to a generative
model~\cite{lewis2021retrievalaugmentedgenerationknowledgeintensivenlp}. RAG reduces hallucination compared with unaugmented generation, but does not eliminate it. Hallucination occurs both when the relevant passage is not retrieved and when the correct passage is retrieved but the model extrapolates beyond it~\cite{https://doi.org/10.1111/jels.12413,hallucination-mitigation-rag}. Beyond factual accuracy, a further problem arises in legal contexts: even a factually correct response does not guarantee that the correct decision procedure was applied to the user's specific facts. Legal eligibility determinations require correct instantiation
of a rule's preconditions in a specific order under specific definitional constraints, and RAG provides no formal guarantee that the model's reasoning follows the legally prescribed framework~\cite{dahl2024largelegal, stanford2025legaldesign}.

\begin{algorithm}[t]
\caption{Purely LLM-based conversational interface using retrieval-augmented generation.}
\label{alg:rag}
\begin{algorithmic}[1]
\Require Knowledge base $KB$, language model $M$
\State $history \gets \emptyset$
\Repeat
  \State $q \gets$ receive free-text utterance from user
  \State $passages \gets \textsc{Retrieve}(KB, q, history)$
  \State $prompt \gets \textsc{Compose}(history, q, passages)$
  \State $response \gets M.\textsc{Generate}(prompt)$
  \State present $response$ to user
  \State $history \gets history \cup \{(q, response)\}$
\Until{$response$ contains a final outcome \textbf{or} user ends session}
\State \Return $\textsc{ExtractOutcome}(response)$
\end{algorithmic}
\end{algorithm}

Algorithm~\ref{alg:rag} formalizes the corresponding RAG interaction loop. The system maintains a running $history$ of past user utterances and model responses, but no structured representation of the underlying administrative process. At each turn, the user's free-text query $q$ is used to retrieve the top-ranked passages from the knowledge base $KB$, with the conversation history available as additional context. The retrieved passages are composed with $history$ and $q$ into a single prompt, which the language model $M$ uses to generate a response; the response is presented to the user and appended to $history$. After each response, the loop terminates if the model has produced a final answer (e.g., an eligibility decision); otherwise, it continues. Crucially, whether the loop terminates and what outcome it returns are both determined by the model or the user.

\subsection{A Hybrid Neuro-Symbolic Architecture}
\label{sec:hybrid}
\begin{figure}
  \centering
\includegraphics[width=.5\textwidth]{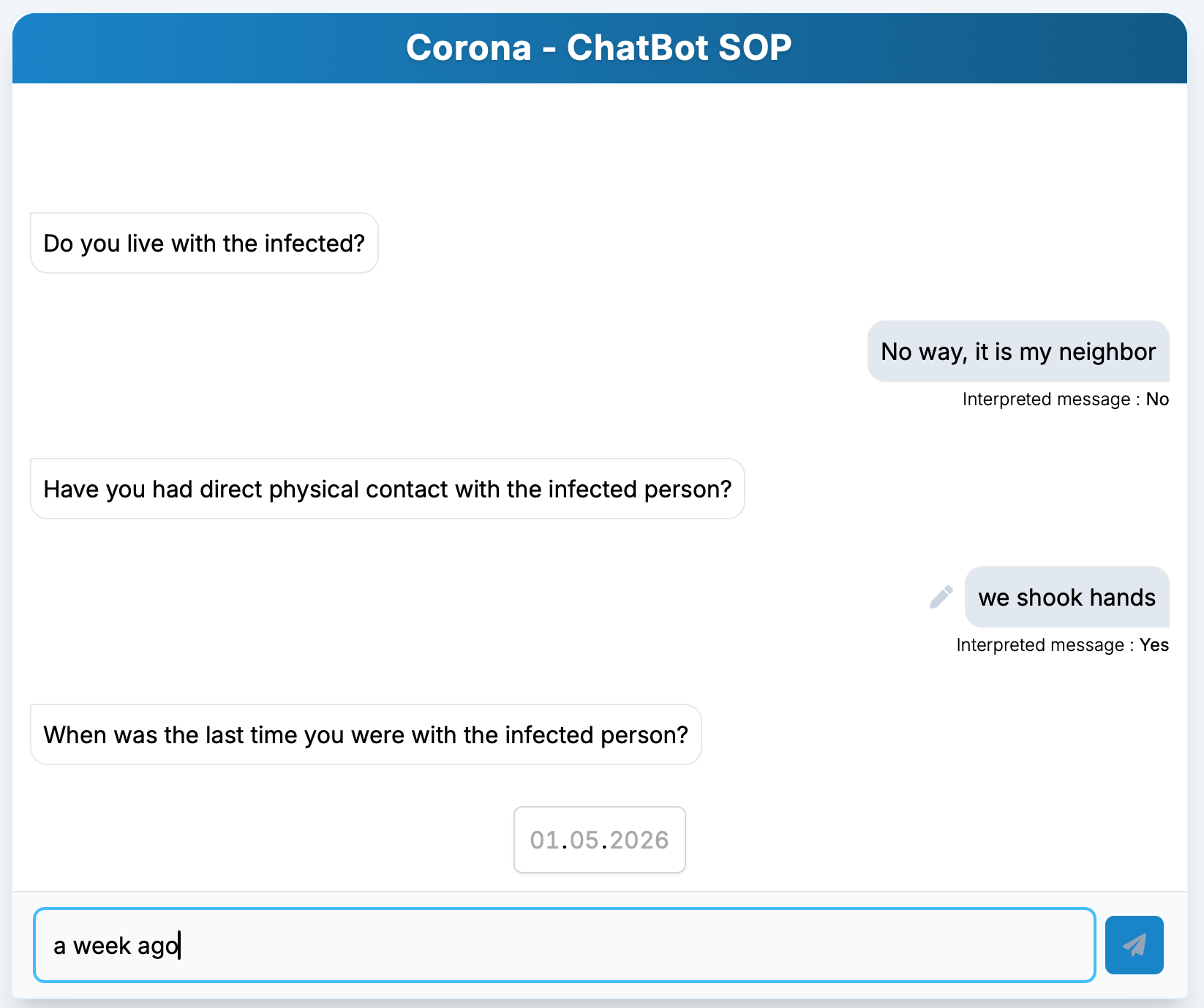}
  \caption{The Chatbot prototype applied to the DCR graph for the Covid-19 guideline (available for test at \url{https://hackathon.dcrgraphs.net/?graphid=2001183}). Note how the citizen has given free text answers and the Chatbot shows how it was interpreted as one of the well-typed possible answers.}
  \Description{The Chatbot prototype applied to the DCR graph for the Covid-19 guideline.}
  \label{fig:CoronaChat}
\end{figure}

The limitations of each approach suggest a principled division of labor: the symbolic component handles the closed-world part of the system, encoding rules, enforcing preconditions, maintaining state, and producing conclusions guaranteed to be consistent with the encoded logic, while the language model handles the open-ended part:
interpreting natural language, resolving ambiguity, and extracting candidate values from free-form text \cite{neurosymbolic-review-springer, garcez2023neurosymbolic}. Fig.~\ref{fig:CoronaChat} shows a run of the prototype implementation of the hybrid architecture. Note that below each answer given by the user is written how the NLP component interpreted the answer, That is, \texttt{"No way, it is my neighbor"} is interpreted as a \texttt{"No"} answer to the question \texttt{"Do you live with the infected?"}.

The symbolic component, $sim$ in Algorithm~\ref{alg:hybrid}, is the same DCR execution engine introduced in Sec.~\ref{sec:API}: it maintains the authoritative process state, and its enabled set evolves with every submitted answer through the same constraint
semantics~\cite{hildebrandt2011declarative}. What is new in the hybrid architecture is the language component (the NLP interpreter 
$N$), which interprets free-text utterances and maps them to structured event--value pairs. In our prototype, $N$ is implemented using an NLP endpoint available in the DCR Active Repository's execution API described in Sec.~\ref{sec:API}, though the architecture itself is agnostic to the choice of NLP component. The two components are connected by a middleware layer that manages per-session state and applies type-level validation to interpretation outputs; neither the symbolic engine nor the NLP component is aware of the other's internal state.

At each conversational turn, the middleware queries the simulation for its current state and selects the next event to present using the same procedure as Algorithm~\ref{alg:symbolic} (line 14). The user may either select from a set of predefined options where applicable (lines 18--19) or respond in free-form natural language. When the user responds in natural language, the utterance is forwarded to the NLP service together with the current event context and all still-open events and their value domains (lines 22-23), providing the constrained hypothesis space of activities that are legally admissible at that point in the process. The service returns whether the utterance addressed the current question, the extracted canonical value, and
optionally inferred values for other open events derivable from the same utterance.
Inferred values are queued and presented to the user for explicit confirmation before the next fresh question is posed (lines 4-12), providing an explicit user-confirmation step before inferred values are committed to the engine. If a value,
entered directly or confirmed from the inferred queue, is accepted, and it is submitted to the simulation engine via the \textsc{Advance} procedure, which recomputes the enabled
and pending sets, and the loop repeats. The user may at any point revisit a previously answered event through the \textsc{Edit} procedure; doing so resets the inferred queue, since values inferred against an earlier answer may no longer be justified once that answer changes.

The primary residual risk concerns the interpretation step. Type-level validation is applied to every returned value (line 29 of Algorithm~\ref{alg:hybrid}), i.e., integers, dates, and choice values are each checked against the event's declared domain, but these checks do not operate at the semantic level. The NLP service does not return a confidence score alongside the extracted value, so no threshold can be applied to reject low-confidence interpretations before they enter the simulation. Instead, the loop falls back to a bounded retry that advances to the next pending event when exhausted (line 32 of Algorithm~\ref{alg:hybrid}). The symbolic engine enforces correctness over all state transitions it receives, but it cannot reject a value that is correctly typed yet incorrectly inferred~\cite{goh2025construct}.
Process-level correctness is therefore preserved; the engine cannot be induced to skip an obligation or reach an unpermitted conclusion, but the process may reach a legitimate terminal state based on a factually incorrect value if the interpretation step misreads the user's intent. However, these errors are mitigated in the prototype by allowing the user to edit a misinterpreted intent, by clicking at the small pencil to the left of the user-given answer shown in Fig.~\ref{fig:CoronaChat}, i.e. ensuring a human-in-the-loop.

\begin{algorithm}[H]
\caption{Hybrid architecture combining a DCR execution engine with an LLM-based interpretation layer.}
\label{alg:hybrid}
\begin{algorithmic}[1]
\Require DCR graph $G$, NLP interpreter $N$
\State $sim \gets \textsc{CreateSimulation}(G)$
\State $history \gets \emptyset$;\quad $inferred \gets \emptyset$
\While{$sim$ has at least one pending event}
  \If{$inferred$ is non-empty}
    \State $(e, v) \gets \textsc{Dequeue}(inferred)$
    \State present $\textsc{Label}(e) = v$ for confirmation
    \If{user confirms}
      \State \textsc{Advance}$(sim, e, v, history)$
    \EndIf
    \State \textbf{continue}
    \Comment{rejected events remain pending}
  \EndIf
  \State $e \gets \textsc{SelectEvent}(sim)$
  \Comment{see Alg.~\ref{alg:symbolic}, lines 6--9}
  \State present $\textsc{Question}(e)$ and $\textsc{Options}(e)$
  \State $input \gets \textsc{UserResponse}()$
  \Comment{option, text, or edit}
  \If{$input$ is edit request for $e_{\text{prev}}$}
    \State \textsc{Edit}$(sim, e_{\text{prev}}, inferred, history)$
    \State \textbf{continue}
  \ElsIf{$input$ is option selection}
    \State $value \gets input.\textit{option}$
  \Else
    \Comment{free-text path}
    \State $ctx \gets sim.\textsc{EnabledNotExecuted}()$
    \State $r \gets N.\textsc{Interpret}(e, input.\textit{text}, ctx)$
    \State $\textsc{Enqueue}(inferred, r.\textit{inferred\_replies})$
    \If{$\neg r.\textit{reply}$ \textbf{or} $r.\textit{value}$ is empty}
      \State request clarification; \textbf{continue}
      \Comment{bounded by $\mathit{MAX\_RETRIES}$}
    \EndIf
    \State $value \gets r.\textit{value}$
  \EndIf
  \If{$\textsc{ValidateType}(e, value)$}
    \State \textsc{Advance}$(sim, e, value, history)$
  \Else
    \State request retry
  \EndIf
\EndWhile
\State \Return $sim.\textsc{Conclusion}()$
\Statex
\Procedure{Advance}{$sim, e, v, history$}
  \State $sim.\textsc{Execute}(e, v)$
  \Comment{recompute enabled/pending/excluded}
  \State $history \gets history \cup \{(e, v)\}$
\EndProcedure
\Statex
\Procedure{Edit}{$sim, e_{\text{prev}}, inferred, history$}
  \If{$e_{\text{prev}}$ is enabled in $sim$}
    \State $inferred \gets \emptyset$
    \State re-present $\textsc{Question}(e_{\text{prev}})$
    \State $v_{\text{new}} \gets \textsc{UserResponse}()$
    \If{$\textsc{ValidateType}(e_{\text{prev}}, v_{\text{new}})$}
      \State \textsc{Advance}$(sim, e_{\text{prev}}, v_{\text{new}}, history)$
    \EndIf
  \EndIf
\EndProcedure
\end{algorithmic}
\end{algorithm}

\section{Discussion and Future Work}

\subsection{DCR Graph Technology Extensions}
The DCR technology and tools, still have some limitations that need to be addressed.

Firstly, the DCR Graphs become difficult to comprehend and manage when they grow to more than 20-30 activities, although some grouping of activities that have rules in common can help reducing the complexity~\cite{10.1007/978-3-031-61057-8_26}. We therefore explore means to support more modularity and reuse in the DCR graph modelling language.

Secondly, the current tools for mapping from text to graphs are based on early language models and only capture the most basic elements of DCR models. In particular, the translations need to deal systematically with the presence of ambiguities in the input texts~\cite{lopez2025ambiguity}. It also completely ignores data and thus cannot identify any data-dependent rules. Moreover, they have been trained for the translation of business process descriptions and not legal texts. We will explore the usage of modern language models for the translation (with human in the loop) of the computational parts of law texts and guidelines into DCR graphs.
Finally, we plan to extend the DCR API to include also end-points for explanations, e.g. for a particular sequence of activities leading to a particular end-state or why an activity is \emph{not} possible to be executed.

\subsection{Testing the Chatbot}
Moving beyond proof-of-concept, the chatbot must be evaluated on legally grounded use cases that reflect realistic interaction flows and decision-making under regulatory constraints. While evaluation on real user interactions would be ideal, privacy and legal considerations prevent their use in our setting. We therefore adopt a controlled simulation approach, constructing synthetic cases that approximate real administrative scenarios and validating them with domain experts (case workers) in the relevant legal areas (e.g., COVID-19 regulation, student disability grants).

On this basis, we aim to develop a legally informed benchmark consisting of a large collection of curated synthetic cases, each reviewed and refined by case workers to ensure fidelity to real-world practice. These cases will support the simulation of complete interaction trajectories, covering both citizen inputs and the corresponding case handling logic. This enables the design of a structured human evaluation protocol targeting (i) legal correctness of the outcome with respect to the underlying rule-based model, (ii) faithfulness in interpreting user-provided information, (iii) clarity and appropriateness of the interaction, and (iv) procedural efficiency.

To complement expert evaluation and enable scalability, we will develop an automated evaluation protocol based on LLM-as-a-judge. Importantly, these automated assessments will be calibrated against expert annotations to ensure alignment with legal reasoning standards. Once validated, the protocol can be applied at scale to systematically analyze performance across the benchmark, identifying strengths, failure modes, and corner cases. This staged approach allows us to defer testing with real case workers in operational settings until sufficient reliability has been established through legally grounded benchmarking.

\subsection{Cognitive Implication for Case Workers}

Case workers play a fundamental role in managing and resolving cases. They hold critical expertise in assessing and addressing several issues emerging while handling new cases, coming from their long-term experience practicing. While AI technologies bring the promise of efficiency, consistency, and cost reduction, as a remedy to overwhelmed bureaucratic systems, there are also considerable threats if such technologies are deployed without appropriate considerations. A prominent threat for knowledge workers, such as case workers, is the potential deskilling~\cite{ferdman2025ai}, i.e., loss of skills, as a result of severe cognitive offloading~\cite{risko2016cognitive} to AI technologies, such as contemporary conversational assistants. \citet{chalkidis2026} surveys the developing literature~\cite{gerlich2025,kosmyna2025, budzyn2025} around this critical topic, manifesting the negative effects of irresponsible AI adoption related to cognitive decline, such as the atrophy of critical thinking. Drawing parallels with the literature, we can imagine that, if not designed appropriately, a highly capable AI conversational agent deployed to assist case workers can, in time, lead to over-reliance on the system's actions and decisions. The case workers may incrementally offload agency over critical tasks, jeopardizing their skills and knowledge due to a lack of personal practice and involvement. We plan to take appropriate consideration in the overall system design to mitigate such threats.

Bearing the cognitive aspects in mind, chatbots have the potential to unify all information retrieval tasks under one user interface with the help of RAG. We aim to develop additional, conservative, extensions of Algorithm~\ref{alg:hybrid}. One such extension, aimed at case workers, will reintegrate the Knowledge base (KB) from Algorithm~\ref{alg:rag} as pairs of DCR Graphs and legal documents. This enables users to retrieve the most relevant DCR Graph and legal text from a given free-text utterance.

\section{Conclusion}
We have presented an approach to build accurate so-called \emph{Conversational Hybrid AI (CHAI)} chatbots by combining a rule based, symbolic AI given as DCR graphs accessible via an API, and an LLM (also accessible via an API). The key idea is to use the symbolic rule model to control the conversation, i.e. the enabled questions and valid types of answers, and then use the LLM to allow the user to provide free natural language texts as input. 
The past experience of modelling rules in guidelines and laws as DCR graphs and the running Conversational Hybrid AI Chatbot prototype demonstrates that the idea is indeed possible in practice.

\begin{acks}
This work is supported by Innovation Fund Denmark.
\end{acks}

\section*{AI Use Statement}
The authors used OpenAI ChatGPT during preparation of this manuscript to assist with outlining the paper, drafting and revising selected passages, improving clarity and structure, and converting author-provided notes into \LaTeX{} prose. The tool was used as a writing and editing aid, not as a source of empirical results, legal authority, or scientific claims. All substantive ideas, architectural choices, examples, and claims were provided, checked, and revised by the authors. The authors independently verified the technical descriptions, legal framing, citations, and consistency with the XHAILe project materials, and take full responsibility for the final content of the paper.

\bibliographystyle{ACM-Reference-Format}
\bibliography{sample-base}

\appendix

\end{document}